\documentclass[reprint,aps,prl,amsmath,amssymb,superscriptaddress]{revtex4-2}

\usepackage{graphicx}
\usepackage{dcolumn}
\usepackage{amsmath}
\usepackage{amssymb}
\usepackage{mathtools}
\usepackage{bbm}
\usepackage{bm}
\usepackage[caption=false]{subfig}
\usepackage{placeins}
\usepackage{amsthm}
\usepackage{natbib}
\usepackage[colorlinks=true,citecolor=blue,linkcolor=blue,urlcolor=blue]{hyperref}

\newcommand{\upe}{\mathrm{e}}   
\newcommand{\upi}{\mathrm{i}}   
\newcommand{\upd}{\mathrm{d}}   

\begin{document}

\title{Dissipative Time Quasicrystals from Multilevel Interference}

\author{Kang Shen}
\affiliation{College of Physical Science and Technology, Central China Normal University, Wuhan 430079, China}

\author{Xiangming Hu}
\email{xmhu@ccnu.edu.cn}
\affiliation{College of Physical Science and Technology, Central China Normal University, Wuhan 430079, China}

\author{Fei Wang}
\email{feiwang@hbut.edu.cn}
\affiliation{School of Science, Hubei University of Technology, Wuhan 430068, China}

\begin{abstract}
	Boundary time crystals exhibit spontaneous breaking of continuous time-translation symmetry through persistent periodic oscillations in driven-dissipative many-body systems. Here, we show that multilevel interference provides a natural route beyond periodic order, enabling dissipative time quasicrystals without externally imposed quasiperiodic driving. We consider a collectively driven-dissipative four-level ensemble with two degenerate excited states and two degenerate ground states. In the thermodynamic limit, the exact mean-field dynamics reduces to an irrational flow on a two-dimensional torus, yielding quasiperiodic order parameters with discrete spectra generated by two incommensurate fundamental frequencies. Vanishing maximal Lyapunov exponents demonstrate that the nonlinear self-consistent dynamics remains nonchaotic. Our results establish a minimal interference-induced mechanism for time-quasiperiodic order and open a route toward higher-dimensional quasiperiodic dynamics in multilevel systems.
\end{abstract}

\maketitle

\emph{Introduction.}---Since the original proposals of time crystals~\cite{Wilczek2012Quantum,Shapere2012Classical}, persistent oscillatory phases have been explored in both periodically driven Floquet systems~\cite{Khemani2016Phase,Else2016Floquet,Yao2017Discrete,Sacha2018Time,RieraCampeny2020Time,Else2020Discrete,Zaletel2023Colloquium} and autonomous driven-dissipative settings~\cite{Iemini2018Boundary,Tucker2018Shattered,Carollo2022Exact,Prazeres2021Boundary,Buca2019Non,Zhu2019Dicke,Cabot2024Nonequilibrium,Nakanishi2023Dissipative,Iemini2024Dynamics,Booker2020Nonstationarity,Buifmmode2019Dissipation,Lledo2019Driven,Seibold2020Dissipative,Hajduifmmode2022Seeding,Hurtado2020Building,Cosme2023Bridging}. A paradigmatic example is provided by boundary time crystals~\cite{Iemini2018Boundary,Prazeres2021Boundary}, where collective dissipation and coherent driving generate robust long-time oscillations without external periodic modulation, thereby spontaneously breaking continuous time-translation symmetry. Existing boundary-time-crystal models typically exhibit periodic motion characterized by a single fundamental frequency~\cite{Iemini2018Boundary,Carollo2022Exact}. A central question is whether time-crystalline order can extend beyond periodic oscillations toward quasiperiodic temporal order. Realizing such behavior without externally imposed quasiperiodic driving or multiple incommensurate drives is highly nontrivial, as previous approaches to temporal quasiperiodicity have largely relied on external multifrequency driving or synthetic constructions~\cite{Autti2018Observation,Giergiel2019Discrete,Zhao2019Floquet,Luo2026Discrete,He2025Experimental,Giergiel2018Time,Pizzi2019Period} rather than intrinsic many-body mechanisms.

Recent studies have shown that multilevel driven-dissipative systems can support a variety of collective phenomena beyond the conventional two-level setting. In Refs.~\cite{Orioli2022Emergent,Sundar2024Squeezing}, interference between different transition pathways was shown to generate an extensive family of collective dark states, emphasizing dissipation-dominated stationary behavior. On the other hand, driven-dissipative systems with richer nonlinear structures were found to exhibit periodic time-crystalline motion, complex temporal oscillations, and chaos~\cite{Prazeres2021Boundary,Solanki2025Chaos}. These developments suggest that multilevel interference may provide a route to intrinsically quasiperiodic collective dynamics in the drive-dominated regime, beyond dissipation-dominated dark states and conventional periodic or chaotic motion.

Here we demonstrate this possibility in a collectively driven-dissipative four-level ensemble, where two transition branches participate in the same coherent drive and collective dissipative channel. When the relative transition weight is irrational, interference between the two collective pathways generates intrinsically quasiperiodic dynamics. In the thermodynamic limit, the exact mean-field dynamics maps onto an irrational flow on a two-dimensional torus, leading to order parameters with two incommensurate fundamental frequencies and discrete quasiperiodic Fourier spectra. Moreover, vanishing maximal Lyapunov exponents demonstrate that the nonlinear self-consistent dynamics remains nonchaotic. We further show that the underlying mechanism naturally generalizes to multilevel systems with rationally independent transition weights, opening a route toward higher-dimensional dissipative time quasicrystals.

\begin{figure*}[htbp]
	\centering
	\includegraphics[scale=0.26]{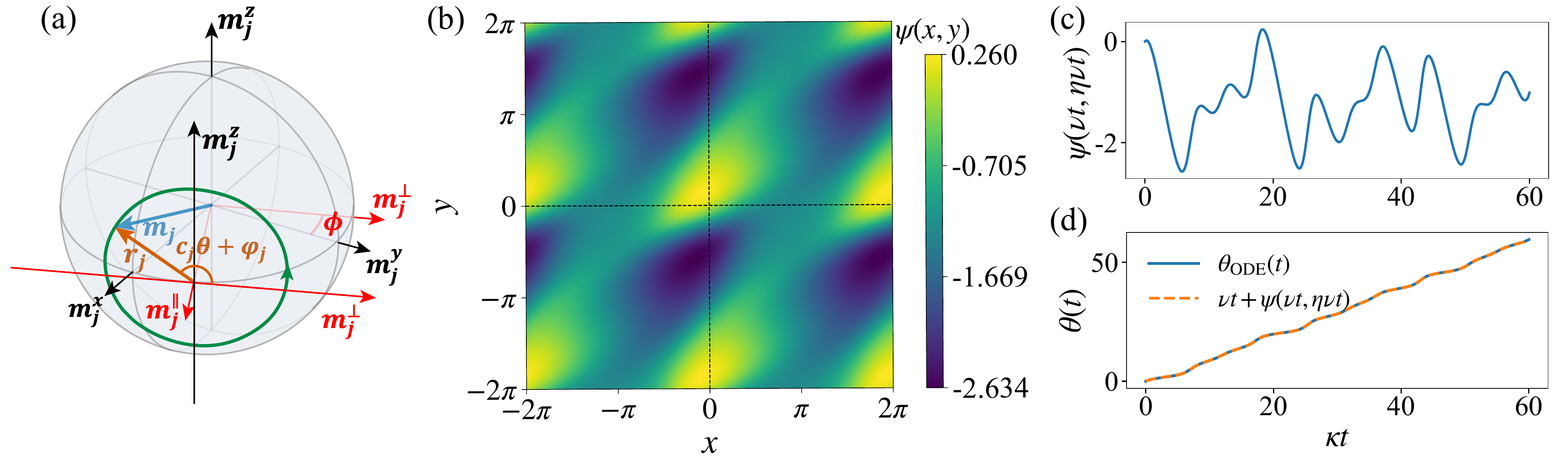}	
	\caption{
		(a) Schematic illustration of the collective Bloch vector $\mathbf{m}_j$ on its Bloch sphere. The green circle shows the trajectory in the plane defined by $\mathbf{e}_\perp$ and $\mathbf{e}_z$, with radius $r_j$ and phase $c_j\theta+\varphi_j$. (b) Color plot of the torus-periodic function $\psi(x,y)$ over four adjacent unit cells to illustrate its two-dimensional periodicity. (c) Quasiperiodic signal $\psi(\nu t,\nu t/\sqrt{2})$ along the irrational flow $(1,1/\sqrt{2})$ on the torus. (d) Comparison of the direct numerical solution of Eq.~\eqref{eq:dottheta} (solid line) and the reconstructed trajectory $\theta(t)=\nu t+\psi(\nu t,\eta\nu t)$ (dashed line), showing excellent agreement and verifying the consistency of the mean drift frequency $\nu$ and the torus correction $\psi(x,y)$. The parameters used are $\omega/\kappa=1.2$, $\eta=1/\sqrt{2}$, $r_1=r_2=1/2$, and $\varphi_1=\varphi_2=-\pi/2$.
	}
	\label{Fig_1}
\end{figure*}

\emph{Collective driven-dissipative four-level model.}---To go beyond the conventional two-level setting of boundary time crystals~\cite{Iemini2018Boundary,Carollo2022Exact}, we consider a collectively driven-dissipative four-level ensemble of $N$ identical atoms. Each atom consists of two degenerate excited states and two degenerate ground states, giving rise to two allowed transition branches. We denote the corresponding collective spin operators by $S_1^\alpha$ and $S_2^\alpha$, with $\alpha=x,y,z$, and the associated collective lowering operators by $S_1^-$ and $S_2^-$. In a rotating frame, the many-body density matrix obeys
\begin{equation}\label{eq:dotrho}
	\frac{\upd \rho}{\upd t} = -\upi \omega \left[S_1^x + \eta S_2^x , \rho\right] + \frac{2\kappa}{N}\mathcal{D}\left[S_1^- + \eta S_2^-\right]\rho.
\end{equation}
Here, $\mathcal{D}[L]\rho=L\rho L^\dagger-\frac{1}{2}\{L^\dagger L,\rho\}$, $\omega$ is the coherent driving strength, $\kappa$ sets the intensive scale of the collective dissipation, and $\eta\in\mathbb{R}$ controls the relative weight of the second transition branch. The coefficient of the first branch has been set to unity by fixing the overall scale, and the two branches are labeled such that the relative amplitude satisfies $|\eta|\leq 1$. The factor $1/N$ in the dissipative term ensures a well-defined thermodynamic limit as $N\to\infty$. Importantly, the two branches enter a common collective jump operator rather than two independent dissipative channels. The resulting cross terms in the dissipator encode interference between the transition pathways, making this four-level ensemble a minimal setting for interference-induced dissipative many-body dynamics.

A possible realization of Eq.~\eqref{eq:dotrho} can be implemented with multilevel atoms coupled to a selected cavity polarization~\cite{Orioli2022Emergent,Sundar2024Squeezing}. Dipole selection rules and polarization configuration isolate two allowed transition branches, while off-branch transitions are suppressed due to vanishing dipole matrix elements or zero overlap with the cavity mode. The corresponding collective dipole operator is $D^- = c_1 S_1^- + c_2 S_2^-$, with coefficients $c_{1,2}$ set by Clebsch--Gordan coefficients and polarization overlaps. By taking the first branch as the reference, one obtains $D^- \propto S_1^- + \eta S_2^-$ with $\eta = c_2/c_1$. Thus, the relative weight $\eta$ emerges intrinsically from the microscopic multilevel structure; in particular, the Clebsch--Gordan coefficients can naturally produce irrational ratios, providing the key ingredient for the quasiperiodic dynamics discussed below.

\emph{Exact mean-field dynamics in the thermodynamic limit.}---As established in Refs.~\cite{Carollo2021Exactness,Carollo2024Applicability,Carollo2022Exact}, mean-field theory becomes exact for all-to-all-coupled many-body systems in the thermodynamic limit. Quantum fluctuations are then suppressed as $N\to\infty$, and the collective dynamics is governed by closed nonlinear equations for macroscopic observables. We therefore analyze Eq.~\eqref{eq:dotrho} within this exact mean-field framework.

We introduce the intensive collective variables
\[
m_j^\alpha=\frac{2}{N}\langle S_j^\alpha\rangle ,
\qquad
j=1,2,
\quad
\alpha=x,y,z,
\]
which remain finite in the thermodynamic limit. Under the mean-field factorization of collective correlators, Eq.~\eqref{eq:dotrho} yields the closed equations of motion
\begin{subequations}
	\begin{align}
		\dot m_j^x &= c_j \kappa \widetilde{m}^x m_j^z, \\
		\dot m_j^y &= -c_j \omega m_j^z + c_j \kappa \widetilde{m}^y m_j^z ,\\
		\dot m_j^z &= c_j \omega m_j^y - c_j \kappa \left(\widetilde{m}^x m_j^x + \widetilde{m}^y m_j^y\right), 
	\end{align}
\end{subequations}
where $c_1=1$, $c_2 = \eta$, and $\widetilde m^{x,y}=m_1^{x,y}+\eta m_2^{x,y}$. The combinations $\widetilde m^{x,y}$ represent the transverse components of the total collective dipole associated with the common transition channel. Their dependence on both branches reflects the interference induced by the shared collective dissipation. Equivalently, with $\mathbf{m}_j=(m_j^x,m_j^y,m_j^z)^{\mathsf T}$, the mean-field equations can be written in the compact precession form
\begin{equation}
	\dot{\mathbf{m}}_j
	=
	c_j\mathbf{\Omega}\times\mathbf{m}_j,
	\qquad
	\mathbf{\Omega}
	=
	\left(
	\omega-\kappa\widetilde m^y,
	\kappa\widetilde m^x,
	0
	\right)^{\mathsf{T}},
\end{equation}
where all quantities are time dependent. Thus, the two collective branches precess about the same self-consistent field $\mathbf{\Omega}$, while their relative angular velocities are controlled by $\eta$.

The precession form immediately reveals the geometry of the mean-field dynamics. Since $\dot{\mathbf {m}}_j$ is perpendicular to $\mathbf{m}_j$, the length $m_j\equiv \lvert\mathbf m_j\rvert$ of each collective Bloch vector is conserved. With the normalization used above, the two radii satisfy $m_1+m_2=1$. Thus, $\mathbf{m}_j$ evolves on a Bloch sphere of fixed radius $m_j$, and the full trajectory is confined to the product manifold $S^2\times S^2$. In addition, the direction of the self-consistent field $\mathbf{\Omega}$ is conserved. Defining $\phi$ as the angle between $\mathbf{\Omega}$ and the $x$ axis,
\[
\tan\phi
=
\frac{\Omega_y}{\Omega_x}
=
\frac{\kappa\widetilde m^x}
{\omega-\kappa\widetilde m^y},
\]
one finds from the equations of motion that $\phi$ is time independent. Consequently, the two Bloch vectors precess about a common axis fixed by the initial condition, whereas the magnitude of $\mathbf{\Omega}$ remains self-consistently determined by their instantaneous positions.

To obtain the explicit trajectory, we introduce a rotated frame in Fig.~\ref{Fig_1}(a). The unit vector $\mathbf{e}_{\parallel}$ is chosen along the conserved direction of $\mathbf{\Omega}$, while $\mathbf{e}_{\perp}$ is chosen such that $\mathbf{e}_{\parallel}\times\mathbf{e}_{\perp}=\mathbf{e}_z$. In this frame, the motion of each Bloch vector is confined to the $\mathbf{e}_\perp$-$\mathbf{e}_z$ plane. It is therefore convenient to regard this plane as a complex plane and write~\cite{Orioli2022Emergent,Sundar2024Squeezing}
\[
m_j^\perp+\upi m_j^z
=
r_j \upe^{\upi(c_j\theta+\varphi_j)} ,
\]
where $r_j$ and $\varphi_j$ are constants fixed by the initial condition. Hence, the full mean-field trajectory is determined once the scalar angle $\theta(t)$ is known. Substituting this parametrization into the self-consistent field yields
\begin{equation}\label{eq:dottheta}
	\dot{\theta}
	=
	\omega_\phi
	-
	\kappa
	\sum_{j=1}^2
	c_j r_j \cos(c_j\theta+\varphi_j),
\end{equation}
where $\omega_\phi=\omega\cos\phi$. Equation~\eqref{eq:dottheta} reduces the nonlinear dynamics on $S^2\times S^2$ to a one-dimensional flow. Depending on the relative strength of the drive term $\omega_\phi$ and the interference-induced dissipative term, two qualitatively distinct regimes arise. In the dissipation-dominated regime, the right-hand side of Eq.~\eqref{eq:dottheta} has zeros and the dynamics relaxes to a fixed point. In the drive-dominated regime, no fixed point exists; $\theta(t)$ increases monotonically, and the two Bloch vectors execute persistent motion on their fixed Bloch spheres.

This structure differs fundamentally from the conventional two-level boundary time crystal. In the latter, there exists a unique global critical point at $\omega=\kappa$. For $\omega<\kappa$, the system relaxes to a stationary phase, whereas for $\omega>\kappa$ it enters a time-crystal phase associated with spontaneous breaking of continuous time-translation symmetry~\cite{Iemini2018Boundary}. In the present multilevel system, however, the interference between the two transition branches qualitatively modifies the critical structure. Destructive interference between the two branches can strongly suppress the effective transverse dipole components $\widetilde{m}^{x,y}$. As a result, the effective drive term is well approximated by $\omega_\phi \simeq \omega$, while the amplitudes $r_j$ can continuously vary between $0$ and $m_j$. The critical condition therefore depends explicitly on the conserved sectors, rather than being determined by a unique global threshold. Different invariant sectors can undergo the transition to persistent motion at different critical points. Nevertheless, one can still define a global critical point $\omega_c$ as the supremum over all sector-dependent thresholds. Remarkably, this global critical point coincides with that of the conventional two-level boundary time crystal, $\omega_c = \kappa$. For $\omega>\omega_c$, all dynamically accessible sectors exhibit persistent motion, implying global spontaneous breaking of continuous time-translation symmetry.

\begin{figure*}[htbp]
	\centering
	\includegraphics[scale=0.3]{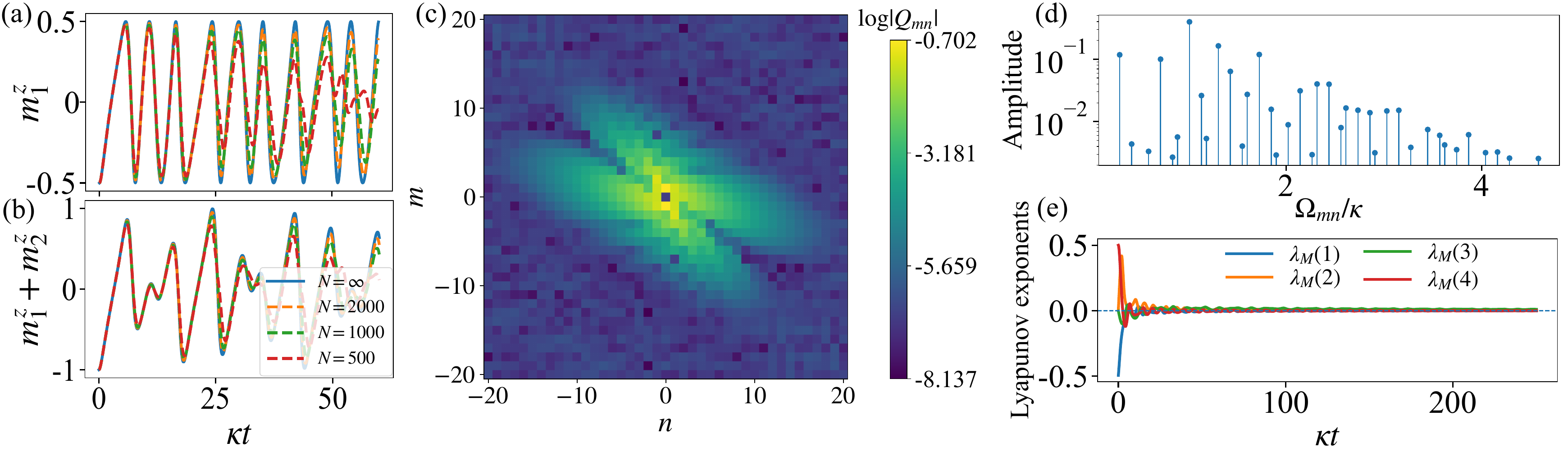}
	\caption{
		(a),(b) Comparison between the thermodynamic-limit mean-field dynamics ($N=\infty$) and finite-$N$ cumulant-expansion results for the order parameters $m_1^z$ and $m_1^z+m_2^z$, respectively. The finite-size trajectories progressively converge toward the thermodynamic-limit solution as the particle number increases. (c) Two-dimensional Fourier amplitudes $\log|Q_{mn}|$ on the $(m,n)$ lattice. (d) One-sided frequency spectrum showing the positions and amplitudes of the dominant quasiperiodic spectral lines at $\Omega_{mn}=(m+\eta n)\nu$; only lines with amplitudes larger than $0.5\%$ of the maximum amplitude are displayed. (e) Time evolution of the maximal Lyapunov exponents for several representative trajectories on $S^2\times S^2$. The initial conditions are $(\mathbf m_1,\mathbf m_2)=(0,0,-1/2,0,0,-1/2)$ for $\lambda_M(1)$, $(0,1/2,0,0,0,1/2)$ for $\lambda_M(2)$, $(1/2,0,0,0,1/2,0)$ for $\lambda_M(3)$, and $(0,0,1/2,0,0,-1/2)$ for $\lambda_M(4)$. In all cases, the maximal Lyapunov exponents decay toward zero, confirming the absence of chaotic dynamics. Parameters are the same as in Fig.~\ref{Fig_1}.
	}	
	\label{Fig_2}
\end{figure*}

\emph{Dissipative time-quasiperiodic dynamics.}---We now focus on the regime $\omega>\kappa$, where the right-hand side of Eq.~\eqref{eq:dottheta} remains positive for all $\theta$. The system is then in the drive-dominated running regime. Taking $\eta$ to be an irrational number satisfying a Diophantine condition~\cite{Arnold1989Mathematical,Broer1990Unfoldings}, the right-hand side of Eq.~\eqref{eq:dottheta} becomes a quasiperiodic function of $\theta$. In this case, the solution can be represented as
\begin{equation}\label{eq:theta_t}
	\theta(t) = \nu t+\psi(\nu t,\eta\nu t),
\end{equation}
where $\nu = \lim_{t\to\infty}\frac{\theta(t)}{t}$ is the mean drift frequency. The function $\psi(x,y)$ is periodic on the two-dimensional torus $\mathbb{T}^2$, satisfying $\psi(x+2\pi,y)=\psi(x,y)$ and $\psi(x,y+2\pi)=\psi(x,y)$. The bounded correction arises by evaluating the torus-periodic function $\psi(x,y)$ along the irrational linear flow $(x,y)=(\nu t,\eta\nu t)$. Hence, for irrational $\eta$, $\psi(\nu t,\eta\nu t)$ is quasiperiodic in time, with fundamental frequencies $\nu$ and $\eta\nu$.

The mean drift frequency $\nu$ can be obtained from the torus average of the inverse phase velocity. To this end, we write the right-hand side of Eq.~\eqref{eq:dottheta} as the restriction $f(\theta)=F(\theta,\eta\theta)$ of a periodic function on $\mathbb T^2$, with
\[
F(x,y)=\omega_\phi-\kappa\left[
r_1\cos(x+\varphi_1)+\eta r_2\cos(y+\varphi_2)
\right].
\]
For the running regime considered here, $F(x,y)>0$ on $\mathbb T^2$. Defining $G(x,y)\equiv1/F(x,y)$, one obtains
\begin{equation}
	\nu^{-1} = G_{00} =\frac{1}{(2\pi)^2}\int_0^{2\pi}\int_0^{2\pi}G(x,y)\,\upd x\,\upd y ,
\end{equation}
where $G_{00}$ is the zeroth Fourier coefficient, equivalently the torus average, of $G(x,y)$. This average can be evaluated analytically in terms of a complete elliptic integral of the first kind; the explicit expression is given in the Supplemental Material.

The torus-periodic correction $\psi(x,y)$ generally has no closed analytic form and is obtained numerically from the corresponding torus equation, derived by substituting Eq.~\eqref{eq:theta_t} into Eq.~\eqref{eq:dottheta}; see the Supplemental Material for details. The solution is computed with periodic boundary conditions on $\mathbb T^2$ and the gauge-fixing condition $\psi(0,0)=0$. Figure~\ref{Fig_1}(b) shows a color plot of $\psi(x,y)$ for representative parameters in the drive-dominated regime. The function is displayed over four neighboring unit cells to make its torus periodicity explicit. Sampling $\psi(x,y)$ along the irrational flow $(x,y)=(\nu t,\nu t/\sqrt{2})$ gives the quasiperiodic correction $\psi(\nu t,\nu t/\sqrt{2})$, shown in Fig.~\ref{Fig_1}(c). Finally, Fig.~\ref{Fig_1}(d) compares the direct numerical solution of Eq.~\eqref{eq:dottheta} with the reconstructed trajectory $\theta(t)=\nu t+\psi(\nu t,\nu t/\sqrt{2})$. Their excellent agreement confirms the consistency of the extracted mean drift frequency $\nu$ and the torus correction $\psi(x,y)$. It is worth emphasizing that $\nu$ and $\psi(x,y)$ play a central role in characterizing the time-quasiperiodic behavior of the order parameters and their Fourier spectrum.

The quasiperiodic structure of the order parameters follows directly from the torus representation of $\theta(t)$. For example, using $m_1^z=r_1\sin(\theta+\varphi_1)$, one obtains
\begin{equation}
	m_1^z(t) = r_1\sin\!\left[ \nu t+\psi(\nu t,\eta\nu t)+\varphi_1 \right].
\end{equation}
Equivalently, $m_1^z(t)$ can be viewed as the restriction of the torus-periodic function $Q(x,y) = r_1\sin\!\left[x+\psi(x,y)+\varphi_1\right]$ along the irrational flow $(x,y)=(\nu t,\eta\nu t)$ on $\mathbb{T}^2$. Consequently, the order parameter exhibits time-quasiperiodic behavior with two fundamental frequencies $\nu$ and $\eta\nu$. To verify this prediction beyond the thermodynamic-limit mean-field theory, we compare the analytic mean-field dynamics with finite-$N$ cumulant-expansion calculations. As shown in Figs.~\ref{Fig_2}(a,b), the finite-$N$ trajectories of $m_1^z$ and $m_1^z+m_2^z$ progressively converge toward the thermodynamic-limit solution as the particle number increases. This agreement confirms both the validity of the exact mean-field description and the quasiperiodic nature of the collective dynamics.

The quasiperiodic nature of the order parameter is further revealed by its Fourier spectrum. Since $Q(x,y)$ is periodic on $\mathbb{T}^2$, it admits the two-dimensional Fourier expansion $Q(x,y) = \sum_{m,n\in\mathbb Z} Q_{mn}\upe^{\upi(mx+ny)}$. Evaluating this expansion along the irrational flow $(x,y)=(\nu t,\eta\nu t)$ gives
\begin{equation}
	m_1^z(t)
	=
	\sum_{m,n\in\mathbb Z}
	Q_{mn}
	\upe^{\upi(m+\eta n)\nu t}.
\end{equation}
Thus, the frequency spectrum consists of discrete lines at $\Omega_{mn}=(m+\eta n)\nu$, which are integer linear combinations of the two fundamental frequencies $\nu$ and $\eta\nu$. Figure~\ref{Fig_2}(c) shows the two-dimensional Fourier amplitudes $\lvert Q_{mn} \rvert$ on the $(m,n)$ lattice. In Fig.~\ref{Fig_2}(d), the same information is projected onto the positive-frequency axis, yielding a one-sided spectrum that displays both the positions and amplitudes of the quasiperiodic spectral lines.

To further characterize the dynamical nature of the quasiperiodic motion, we analyze the maximal Lyapunov exponent, which quantifies the asymptotic rate of exponential separation between nearby trajectories and serves as a standard diagnostic of chaos~\cite{Lerose2020Bridging,Thompson2002Nonlinear,Vulpiani2010Chaos,Benettin1976Kolmogorov,Benettin1980Lyapunov,Benettin1980Lyapunov2}. Since the mean-field dynamics is constrained to the manifold $S^2\times S^2$ by the conservation of $|\mathbf m_j|$, only tangential perturbations on the Bloch spheres are considered, while radial directions are excluded from the stability analysis. Figure~\ref{Fig_2}(e) shows the time evolution of the maximal Lyapunov exponents for several representative trajectories. In all cases, the exponents decay toward zero at long times rather than approaching positive values. This demonstrates the absence of chaotic dynamics and indicates that the motion remains nonchaotic despite the nonlinear self-consistent evolution. Combined with the discrete frequency spectrum shown in Fig.~\ref{Fig_2}(d), the vanishing maximal Lyapunov exponents confirm that the dynamics is quasiperiodic rather than chaotic.

Taken together, these results establish a robust form of dissipative time-quasiperiodic dynamics in the thermodynamic limit. The collective motion is characterized by two incommensurate fundamental frequencies, discrete quasiperiodic spectral lines, and vanishing maximal Lyapunov exponents, demonstrating that the dynamics remains nonchaotic despite the nonlinear self-consistent evolution. Unlike conventional boundary time crystals, where the long-time dynamics is purely periodic, the present multilevel system supports intrinsically quasiperiodic collective motion generated by interference between different transition branches. This provides a minimal driven-dissipative setting for realizing dissipative time quasicrystals without external quasiperiodic modulation.

\emph{Generalizations to multilevel systems.}---The mechanism underlying the present dissipative time-quasiperiodic dynamics naturally extends beyond the minimal four-level setting considered here. In a generic multilevel driven-dissipative ensemble, different transition branches can enter the same coherent drive and collective dissipative channel with distinct relative weights. When these weights are rationally independent, the resulting dynamics can involve multiple incommensurate frequencies and therefore generate higher-dimensional quasiperiodic motion on $\mathbb T^d$. 

In this generalized setting, the order parameters are expected to exhibit multidimensional quasiperiodic spectra consisting of discrete frequency combinations of several fundamental modes. Importantly, the physical origin of the quasiperiodicity remains the same as in the four-level model: it arises intrinsically from interference between different collective transition pathways rather than from externally imposed quasiperiodic driving. These multilevel systems therefore provide a natural route toward higher-dimensional dissipative time quasicrystals in driven-dissipative many-body systems.

\emph{Acknowledgments.}---This work was supported by the National Natural Science Foundation of China (Grant Nos. 12274164, 61875067, and 12375011).

\emph{Data availability.}---The data that support the findings of this article is available in Zenodo at~\cite{Shen2026Dataset}.


%

\setcounter{secnumdepth}{2}
\clearpage
\setcounter{equation}{0}
\setcounter{figure}{0}
\setcounter{table}{0}
\setcounter{section}{0}

\makeatletter
\renewcommand{\theequation}{S\arabic{equation}}
\renewcommand{\thefigure}{S\arabic{figure}}
\renewcommand{\bibnumfmt}[1]{[S#1]}
\renewcommand\thesection{\Roman{section}}

\onecolumngrid

\begin{center}
	\Large \textbf{Supplemental Material for ``Dissipative Time Quasicrystals from Multilevel Interference''}
	
	\vspace{0.5cm}
	
	\normalsize
	Kang Shen$^{1}$, Xiangming Hu$^{1,*}$, and Fei Wang$^{2,\dag}$
	
	\vspace{0.2cm}
	
	\textit{$^1$College of Physical Science and Technology, Central China Normal University, Wuhan 430079, China} \\
	\textit{$^2$School of Science, Hubei University of Technology, Wuhan 430068, China}
\end{center}

\vspace{8mm}


The Supplemental Material is organized as follows. In Sec.~I, we derive the mean-field dynamics from the master equation, identify the constants of motion, and reduce the collective Bloch-vector dynamics to a scalar phase equation. This reduction also clarifies how multilevel interference modifies the running regime compared with the standard two-level boundary time crystal. In Sec.~II, we develop the quasiperiodic description of the running phase, derive the form \(\theta(t)=\nu t+\psi(\nu t,\eta\nu t)\), and obtain the associated frequency spectrum and the analytic expression for the mean drift frequency.

\section{Mean-Field Dynamics and Constants of Motion}
\subsection{Bloch-vector equations}
Starting from the master equation in Eq.~(1) of the main text,
\begin{equation}
	\dot{\rho} = -\upi\omega [S_1^x+\eta S_2^x,\rho] + \frac{2\kappa}{N} \mathcal D[S_1^-+\eta S_2^-]\rho,
\end{equation}
we first derive the equations of motion for the collective spin operators. For an arbitrary operator $O$, its time evolution is given by
\[
\frac{d}{dt}\langle O\rangle = -\upi\omega \langle [O,S_1^x+\eta S_2^x]\rangle + \frac{2\kappa}{N} \left\langle L^\dagger O L - \frac{1}{2}\{L^\dagger L,O\}\right\rangle,
\]
where $L=S_1^-+\eta S_2^-$. Using the angular-momentum commutation relations $[S_i^\alpha,S_j^\beta] = \upi\delta_{ij}\epsilon_{\alpha\beta\gamma}S_j^\gamma$, one obtains the exact equations of motion
\begin{align}
	\dot{S}_j^x &= c_j\frac{\kappa}{N} \left(\{\widetilde{S}^x,S_j^z\} - c_jS_j^x\right),
	\\
	\dot{S}_j^y &= -c_j\omega S_j^z + c_j\frac{\kappa}{N} \left(\{\widetilde{S}^y,S_j^z\} - c_jS_j^y\right),
	\\
	\dot{S}_j^z &= c_j\omega S_j^y - 2c_j\frac{\kappa}{N} \left(\widetilde{S}^xS_j^x + \widetilde{S}^yS_j^y + c_jS_j^z\right),
\end{align}
where \(j=1,2\), \(c_1=1\), \(c_2=\eta\), and
\[
\widetilde S^\alpha = S_1^\alpha+\eta S_2^\alpha.
\]
Here and below, operator products in the equations of motion are understood inside expectation values.

We then introduce the intensive collective Bloch variables
\[
m_j^\alpha = \frac{2}{N}\langle S_j^\alpha\rangle,
\]
which remain finite in the thermodynamic limit. Since the dissipative coupling is scaled as \(1/N\), the terms involving products of collective spin operators contribute at the same order as the coherent driving. The mean-field limit is obtained by factorizing two-body correlations as
\begin{equation}
	\langle S_i^\alpha S_j^\beta\rangle \simeq \langle S_i^\alpha\rangle \langle S_j^\beta\rangle,
\end{equation}
while subleading terms proportional to $1/N$ are neglected. Equivalently, the connected correlations,
\[
\langle S_i^\alpha S_j^\beta\rangle - \langle S_i^\alpha\rangle \langle S_j^\beta\rangle,
\]
give corrections that vanish after normalization by the collective spin length in the thermodynamic limit~\cite{Carollo2021ExactnessSM,Carollo2022ExactSM,Carollo2024ApplicabilitySM}.

Substituting
\[
\widetilde m^\alpha = m_1^\alpha+\eta m_2^\alpha
\]
into the normalized equations, we obtain the closed mean-field dynamics
\begin{align}
	\dot m_j^x &= c_j\kappa \widetilde m^x m_j^z,
	\\
	\dot m_j^y &= -c_j\omega m_j^z + c_j\kappa \widetilde m^y m_j^z,
	\\
	\dot m_j^z &= c_j\omega m_j^y - c_j\kappa \left(\widetilde m^x m_j^x + \widetilde m^y m_j^y\right).
\end{align}
The terms proportional to $c_j^2/N$ in the exact operator equations drop out in this limit and therefore do not appear in the mean-field equations.

These equations can be recast into the compact precession form
\begin{equation}
	\dot{\mathbf m}_j = c_j\,\mathbf\Omega\times \mathbf m_j,
\end{equation}
with the self-consistent effective field
\begin{equation}
	\mathbf\Omega = \left(\omega-\kappa \widetilde m^y,\kappa \widetilde m^x,0\right)^{\mathsf{T}}.
\end{equation}
This form shows that each Bloch vector precesses around the same instantaneous field $\mathbf{\Omega}$, but with a branch-dependent rate set by $c_j$. For all-to-all coupled collective systems, such a mean-field description becomes exact in the thermodynamic limit $N\to\infty$, where normalized connected correlations are suppressed by powers of $1/N$.

\subsection{Constants of motion}
The precession structure of the mean-field equations immediately implies the existence of several conserved quantities. These conserved quantities strongly constrain the dynamics and reduce the effective phase-space dimension.

First, the length of each Bloch vector is conserved. Taking the time derivative of
\begin{equation}
	|\mathbf m_j|^2 = (m_j^x)^2+(m_j^y)^2+(m_j^z)^2 ,
\end{equation}
one obtains
\begin{align}
	\frac{d}{dt}(m_j^x)^2 &= 2c_j\kappa \widetilde m^x m_j^x m_j^z，
	\\
	\frac{d}{dt}(m_j^y)^2 &= -2c_j\omega m_j^y m_j^z + 2c_j\kappa \widetilde m^y m_j^y m_j^z,
	\\
	\frac{d}{dt}(m_j^z)^2 &= 2c_j\omega m_j^y m_j^z - 2c_j\kappa \widetilde m^x m_j^x m_j^z - 2c_j\kappa \widetilde m^y m_j^y m_j^z .
\end{align}
Summing the three contributions gives
\[
\frac{d}{dt}|\mathbf m_j|^2 = 0 ,
\]
which implies
\begin{equation}
	|\mathbf m_j| = \mathrm{const}.
\end{equation}

Therefore, the dynamics of each branch is restricted to the surface of a Bloch sphere with fixed radius. Since the total spin length is conserved, the two radii satisfy
\begin{equation}
	|\mathbf m_1| + |\mathbf m_2| = 1 .
\end{equation}

A second conserved quantity follows from the dynamics of the effective
field
\begin{equation}
	\mathbf\Omega(t) = (\omega-\kappa \widetilde m^y,\kappa \widetilde m^x,0)^{\mathsf{T}}.
\end{equation}
Taking the time derivative yields
\begin{align}
	\frac{d}{dt}\mathbf\Omega(t) = \kappa(-\dot{\widetilde m}^y,\dot{\widetilde m}^x,0)^{\mathsf{T}}.
\end{align}
Using $\dot{\mathbf m}_j = c_j\mathbf{\Omega} \times \mathbf{m}_j$, one finds
\begin{equation}
	\frac{d}{dt}\mathbf\Omega(t) = \kappa (0,0,1)^{\mathsf T} \times \left[\sum_{j=1}^2 c_j^2 \mathbf\Omega(t)\times \mathbf m_j\right] = \kappa \sum_{j=1}^2 c_j^2 m_j^z \mathbf\Omega(t),
\end{equation}
where the vector triple-product identity has been used, together with the fact that $\Omega_z=0$.

Therefore, the time derivative of $\mathbf{\Omega}$ is always parallel to $\mathbf{\Omega}$ itself. As a consequence, the direction of the effective field remains invariant during the dynamics. Equivalently,
\begin{equation}
	\frac{\kappa \widetilde m^x}{\omega-\kappa \widetilde m^y} = \mathrm{const},
\end{equation}
which determines a conserved angle
\begin{equation}
	\tan\phi = \frac{\kappa \widetilde m^x}{\omega-\kappa \widetilde m^y}.
\end{equation}
Geometrically, this means that all Bloch vectors precess around a common fixed direction in spin space, while the magnitude of the effective field may still evolve dynamically.

\subsection{Reduction to the phase equation}
Since the direction of the effective field
\begin{equation}
	\mathbf\Omega(t) = (\omega-\kappa \widetilde m^y,\kappa \widetilde m^x,0)^{\mathsf{T}}
\end{equation}
is conserved, it is convenient to introduce a fixed orthonormal basis
\((\mathbf e_\parallel,\mathbf e_\perp,\mathbf e_z)\), where
\begin{equation}
	\mathbf e_\parallel = \frac{\mathbf\Omega(t)}{|\mathbf\Omega(t)|} = (\cos\phi,\sin\phi,0)^{\mathsf{T}},
\end{equation}
and
\begin{equation}
	\mathbf e_\perp = (-\sin\phi,\cos\phi,0)^{\mathsf T}.
\end{equation}
Here the conserved angle $\phi$ is determined by
\begin{equation}
	\cos\phi = \frac{\omega-\kappa \widetilde m^y}{\sqrt{(\omega-\kappa \widetilde m^y)^2+\kappa^2(\widetilde m^x)^2}},\qquad \sin\phi = \frac{\kappa \widetilde m^x}{\sqrt{(\omega-\kappa \widetilde m^y)^2+\kappa^2(\widetilde m^x)^2}} .
\end{equation}

In this basis, each Bloch vector can be decomposed as
\begin{equation}
	\mathbf m_j(t) = a_j \mathbf e_\parallel + r_j\cos(c_j\theta(t)+\varphi_j)\mathbf e_\perp + r_j\sin(c_j\theta(t)+\varphi_j)\mathbf e_z ,
\end{equation}
where \(a_j\), \(r_j\), and \(\varphi_j\) are determined by the initial condition. Explicitly,
\begin{equation}
	a_j
	=
	\mathbf m_j(0)\cdot \mathbf e_\parallel
	=
	m_j^x(0)\cos\phi+m_j^y(0)\sin\phi ,
\end{equation}
and
\begin{equation}
	r_j
	=
	\sqrt{|\mathbf m_j|^2-a_j^2}.
\end{equation}
The initial phase \(\varphi_j\) is determined by
\begin{equation}
	\cos\varphi_j
	=
	\frac{
		m_j^y(0)\cos\phi
		-
		m_j^x(0)\sin\phi
	}{r_j},
	\qquad
	\sin\varphi_j
	=
	\frac{m_j^z(0)}{r_j}.
\end{equation}

The precession equation
\begin{equation}
	\dot{\mathbf m}_j = c_j\mathbf\Omega\times \mathbf m_j
\end{equation}
then implies that all branches are governed by a single phase variable \(\theta(t)\), whose instantaneous velocity is the magnitude of the effective field,
\begin{equation}
	\dot\theta(t) = |\mathbf\Omega(t)| .
\end{equation}

To obtain the explicit form of \(|\mathbf\Omega(t)|\), we introduce the vector perpendicular to $\mathbf\Omega(t)$,
\begin{equation}
	\mathbf\Omega_\perp(t) = (-\kappa \widetilde m^x, \omega-\kappa \widetilde m^y, 0)^{\mathsf{T}}.
\end{equation}
By construction,
\begin{equation}
	\mathbf\Omega_\perp(t)\cdot \mathbf\Omega(t)=0, \qquad |\mathbf\Omega_\perp(t)|=|\mathbf\Omega(t)|.
\end{equation}
Since the direction of \(\mathbf\Omega(t)\) is conserved, the direction of \(\mathbf\Omega_\perp(t)\) is also conserved. Therefore,
\begin{equation}
	\mathbf\Omega_\perp(t) = |\mathbf\Omega(t)|\,\mathbf e_\perp .
\end{equation}
Taking the scalar product with \(\mathbf e_\perp\), we obtain
\begin{equation}
	|\mathbf\Omega(t)| = \mathbf\Omega_\perp(t)\cdot\mathbf e_\perp = \omega\,\mathbf e_y\cdot\mathbf e_\perp - \kappa \sum_{j=1}^2 c_j (m_j^x,m_j^y,0)^{\mathsf T}\cdot\mathbf e_\perp .
\end{equation}
Using
\begin{equation}
	\mathbf e_y\cdot\mathbf e_\perp=\cos\phi
\end{equation}
and the decomposition
\begin{equation}
	\mathbf m_j(t)
	=
	a_j\mathbf e_\parallel
	+
	r_j\cos(c_j\theta(t)+\varphi_j)\mathbf e_\perp
	+
	r_j\sin(c_j\theta(t)+\varphi_j)\mathbf e_z ,
\end{equation}
we have
\begin{equation}
	(m_j^x,m_j^y,0)^{\mathsf T}\cdot\mathbf e_\perp
	=
	r_j\cos(c_j\theta(t)+\varphi_j).
\end{equation}
Hence
\begin{equation}
	|\mathbf\Omega(t)|
	=
	\omega\cos\phi
	-
	\kappa
	\sum_{j=1}^2
	c_j r_j
	\cos(c_j\theta(t)+\varphi_j).
\end{equation}
Since \(\dot\theta(t)=|\mathbf\Omega(t)|\), the scalar phase equation is

\begin{equation}
	\boxed{
		\dot\theta(t) = \omega\cos\phi -\kappa \sum_{j=1}^2 c_j r_j\cos(c_j\theta(t)+\varphi_j)
	}.
\end{equation}

For the present two-branch model, \(c_1=1\) and \(c_2=\eta\), giving
\begin{equation}
	\dot\theta(t)
	=
	\omega\cos\phi
	-
	\kappa
	\left[
	r_1\cos(\theta+\varphi_1)
	+
	\eta r_2\cos(\eta\theta+\varphi_2)
	\right].
\end{equation}
This equation is the central reduced description of the mean-field motion. For irrational \(\eta\), the two arguments \(\theta\) and \(\eta\theta\) wind quasiperiodically on a two-dimensional torus.

\subsection{Two-level limit}
The conventional two-level boundary time-crystal model is recovered by keeping only one collective Bloch vector. In this case, the phase equation reduces to
\begin{equation}
	\dot\theta = \omega\cos\phi - \kappa r\cos(\theta+\varphi).
\end{equation}
A running solution exists when the right-hand side does not vanish, namely when
\begin{equation}
	\omega\cos\phi>\kappa r .
\end{equation}
To determine the corresponding critical point, we express \(r\) in terms of the initial Bloch vector \(\mathbf m(0)=(x,y,z)\). Since
\begin{equation}
	\cos\phi = \frac{\omega-\kappa y}{\sqrt{(\omega-\kappa y)^2+\kappa^2 x^2}}, \qquad \sin\phi = \frac{\kappa x}{\sqrt{(\omega-\kappa y)^2+\kappa^2 x^2}},
\end{equation}
and \(|\mathbf m|=1\), one has
\begin{equation}
	r^2 = 1-(x\cos\phi+y\sin\phi)^2 = \cos^2\phi + \frac{(\kappa^2-\omega^2)x^2}{(\omega-\kappa y)^2+\kappa^2x^2}.
\end{equation}
Therefore,
\begin{equation}
	r^2-\cos^2\phi = \frac{(\kappa^2-\omega^2)x^2}{(\omega-\kappa y)^2+\kappa^2x^2}.
\end{equation}
It follows that
\begin{equation}
	r\geq \cos\phi ,\qquad \kappa\geq \omega ,
\end{equation}
whereas
\begin{equation}
	r<\cos\phi ,\qquad \kappa<\omega ,
\end{equation}
for generic \(x\neq 0\). Consequently, for \(\omega>\kappa\),
\begin{equation}
	\omega\cos\phi > \kappa\cos\phi > \kappa r ,
\end{equation}
so the phase velocity remains positive and the phase keeps running. By contrast, for \(\omega<\kappa\), one has \(r > \cos\phi\), and the phase equation can admit fixed points. Hence the transition between the pinned and running regimes occurs at
\begin{equation}
	\omega_c=\kappa .
\end{equation}

\subsection{Interference-induced modification of the running regime}
The four-level interfering system differs qualitatively from the standard two-level case. To illustrate this difference, we consider the family of initial states
\begin{equation}
	\mathbf m_1(0) = \left(m_1\sqrt{1-\epsilon^2}, 0, m_1\epsilon \right),
	\qquad
	\mathbf m_2(0) = \left(-m_2\sqrt{1-\epsilon^2}, 0, m_2\epsilon \right),
\end{equation}
with \(0\leq \epsilon\leq 1\) and
\begin{equation}
	\frac{m_1}{m_2}=\eta .
\end{equation}
Since \(m_1+m_2=1\), this gives
\begin{equation}
	m_1=\frac{\eta}{1+\eta},
	\qquad
	m_2=\frac{1}{1+\eta}.
\end{equation}
For this choice of initial condition, the transverse components cancel in the collective field,
\begin{equation}
	\widetilde m^x(0) = m_1^x(0)+\eta m_2^x(0) = 0,
	\qquad
	\widetilde m^y(0)=0 .
\end{equation}
Therefore, the effective field initially points along the \(x\) direction, so that
\begin{equation}
	\cos\phi=1 .
\end{equation}
The transverse radii are
\begin{equation}
	r_1=\epsilon m_1,
	\qquad
	r_2=\epsilon m_2 .
\end{equation}
The reduced phase equation then becomes
\begin{equation}
	\dot\theta = \omega - \kappa \left[ r_1\cos(\theta+\varphi_1)+ \eta r_2\cos(\eta\theta+\varphi_2)\right].
\end{equation}
Since
\begin{equation}
	r_1+\eta r_2=\epsilon(m_1+\eta m_2)=\frac{2\epsilon\eta}{1+\eta},
\end{equation}
a sufficient condition for the running regime is
\begin{equation}
	\omega>\kappa(r_1+\eta r_2)=2\epsilon\frac{\eta}{1+\eta}\kappa .
\end{equation}
Thus, for this family of initial states, the running threshold depends on \(\epsilon\). This is in sharp contrast to the standard two-level case, where the transition occurs at the global critical point
\begin{equation}
	\omega_c=\kappa .
\end{equation}

This initial-state-dependent threshold is a direct consequence of interference between different transition channels. In the four-level system, the collective transverse components entering the self-consistent field,
\begin{equation}
	\widetilde m^\alpha = m_1^\alpha+\eta m_2^\alpha ,
	\qquad
	\alpha=x,y ,
\end{equation}
can be partially or completely canceled by destructive interference between the two Bloch vectors. As a result, the self-consistent field felt by the spins can be strongly reduced even when the individual Bloch
vectors remain finite.

Nevertheless, a global sufficient condition for the running regime can still be obtained from the self-consistent field. If the \(x\)-component of the field,
\begin{equation}
	\Omega_x = \omega-\kappa\widetilde m^y ,
\end{equation}
remains strictly positive for all times, then the effective field never vanishes and the system cannot reach a stationary state. Since
\begin{align}
	\kappa\widetilde m^y = \kappa(m_1^y+\eta m_2^y)\leq \kappa(m_1+\eta m_2)=\kappa[m_1+\eta(1-m_1)]\leq\kappa ,
\end{align}
one has
\begin{equation}
	\Omega_x>0 \qquad \text{for all times if} \qquad \omega>\kappa .
\end{equation}
Therefore, \(\omega>\kappa\) provides a global driving-dominated running condition for the interfering four-level system, even though particular families of initial states can enter the running regime at smaller values of \(\omega\).

\section{Quasiperiodic Dynamics}
In the running regime, the reduced phase equation takes the form
\begin{equation}
	\dot\theta = F(\theta,\eta\theta),
\end{equation}
where
\begin{equation}
	F(x,y) = \omega_\phi - A_1\cos(x+\varphi_1) - A_2\cos(y+\varphi_2).
\end{equation}
Here
\begin{equation}
	\omega_\phi=\omega\cos\phi, \qquad A_1=\kappa r_1, \qquad A_2=\kappa\eta r_2 .
\end{equation}
For irrational \(\eta\), the dynamics is quasiperiodically driven by the
two incommensurate phases \(\theta\) and \(\eta\theta\).

\subsection{Mean drift frequency}

We define $G(x,y)=\frac{1}{F(x,y)}$. Since $\dot\theta = F(\theta,\eta\theta)$, one has
\begin{equation}
	\frac{dt}{d\theta} = G(\theta,\eta\theta).
\end{equation}
Integrating from \(0\) to \(\theta\),
\begin{equation}
	t = \int_0^\theta G(s,\eta s)\,ds .
\end{equation}

The mean drift frequency is defined by $\nu = \lim_{t\to\infty} \frac{\theta(t)}{t}$. Equivalently,
\begin{equation}
	\nu^{-1}
	=
	\lim_{\theta\to\infty}
	\frac{t}{\theta}
	=
	\lim_{\theta\to\infty}
	\frac1\theta
	\int_0^\theta
	G(s,\eta s)\,ds .
\end{equation}
Writing
\begin{equation}
	G(x,y)
	=
	\sum_{m,n\in\mathbb Z}
	G_{mn}e^{\upi(mx+ny)},
\end{equation}
where
\begin{equation}
	G_{mn}
	=
	\frac{1}{(2\pi)^2}
	\int_0^{2\pi}\int_0^{2\pi}
	G(x,y)e^{-\upi(mx+ny)}\,dx\,dy ,
\end{equation}
one has
\begin{equation}
	\nu^{-1} =
	\lim_{\theta\to\infty}
	\frac1\theta
	\int_0^\theta
	G(s,\eta s)\,ds =
	\lim_{\theta\to\infty}
	\frac1\theta
	\int_0^\theta
	\sum_{m,n\in\mathbb Z}
	G_{mn}e^{\upi(m+n\eta)s}\,ds .
\end{equation}
For irrational \(\eta\), \(m+n\eta\neq0\) for all
\((m,n)\neq(0,0)\). Hence all oscillatory Fourier modes vanish after the
long-time average:
\begin{equation}
	\lim_{\theta\to\infty}
	\frac1\theta
	\int_0^\theta
	e^{\upi(m+n\eta)s}\,ds
	=
	0 ,
	\qquad
	(m,n)\neq(0,0).
\end{equation}
Only the zero mode remains, giving
\begin{equation}
	\nu^{-1}
	=
	G_{00}
	=
	\frac{1}{(2\pi)^2}
	\int_0^{2\pi}\int_0^{2\pi}
	G(x,y)\,dx\,dy .
\end{equation}
Hence
\begin{equation}
	\boxed{
		\nu
		=
		\left[
		\frac{1}{(2\pi)^2}
		\int_0^{2\pi}\int_0^{2\pi}
		\frac{dx\,dy}
		{
			\omega_\phi
			-
			A_1\cos(x+\varphi_1)
			-
			A_2\cos(y+\varphi_2)
		}
		\right]^{-1}.
	}
\end{equation}
Since the torus average is invariant under translations, \(\varphi_1\) and \(\varphi_2\) do not affect the value of \(\nu\).

\subsection{Construction of the quasiperiodic solution}
We now construct the quasiperiodic form of the phase solution. Let
\begin{equation}
	G(x,y) = \nu^{-1} + \sum_{(m,n)\neq(0,0)} G_{mn}e^{\upi(mx+ny)} .
\end{equation}
To solve the nonzero Fourier modes, we introduce \(U(x,y)\) through
\begin{equation}
	(\partial_x+\eta\partial_y)U(x,y) = G(x,y)-\nu^{-1}.
\end{equation}
Formally,
\begin{equation}
	U(x,y) = \sum_{(m,n)\neq(0,0)} \frac{G_{mn}}{\upi(m+n\eta)} e^{\upi(mx+ny)} .
\end{equation}

Here the irrationality of \(\eta\) ensures that
\begin{equation}
	m+n\eta\neq0
\end{equation}
for all \((m,n)\neq(0,0)\). To control the small denominators, we further assume that \(\eta\) satisfies a Diophantine condition, namely there exist constants \(C>0\) and \(\tau>0\) such that
\begin{equation}
	|m+n\eta| \geq \frac{C}{(|m|+|n|)^\tau}, \qquad (m,n)\neq(0,0).
\end{equation}
Under this condition, the Fourier series for \(U(x,y)\) is well defined for sufficiently smooth \(G(x,y)\), and \(U(x,y)\) is a periodic function on the two-dimensional torus.

Substituting \(x=\theta\) and \(y=\eta\theta\), we obtain
\begin{equation}
	\frac{dt}{d\theta} = \nu^{-1} + \frac{d}{d\theta}U(\theta,\eta\theta).
\end{equation}
Integrating over \(\theta\) gives
\begin{equation}
	t = \nu^{-1}\theta + U(\theta,\eta\theta) - U(0,0).
\end{equation}
Since a constant shift of \(U\) has no physical effect, we choose the gauge \(U(0,0)=0\), yielding
\begin{equation}
	t = \nu^{-1}\theta + U(\theta,\eta\theta).
\end{equation}
Equivalently,
\begin{equation}
	\theta(t) = \nu t - \nu U(\theta(t),\eta\theta(t)).
\end{equation}
This expression already shows that the phase is a uniform drift plus a bounded modulation. Since \(U\) is bounded and periodic, one has
\begin{equation}
	\theta(t)\sim \nu t
\end{equation}
at long times. Therefore, it is natural to seek the solution in the form
\begin{equation}
	\theta(t) = \nu t+\psi(\nu t,\eta\nu t),
\end{equation}
where \(\psi(x,y)\) is expected to be a periodic function on the two-dimensional torus.

We now justify the periodicity of \(\psi(x,y)\). Let
\begin{equation}
	x=\nu t, \qquad y=\eta\nu t,
\end{equation}
and write
\begin{equation}
	\theta=x+\psi(x,y).
\end{equation}
Substituting this into
\begin{equation}
	t = \nu^{-1}\theta + U(\theta,\eta\theta)
\end{equation}
gives
\begin{equation}
	0 = \psi(x,y) + \nu U(x+\psi(x,y),y+\eta\psi(x,y)).
\end{equation}
Thus, for each \((x,y)\), \(\psi(x,y)\) is determined as the zero of
\begin{equation}
	P_{x,y}(s) = s+\nu U(x+s,y+\eta s).
\end{equation}
That is,
\begin{equation}
	P_{x,y}(\psi(x,y))=0.
\end{equation}

The periodicity follows directly from the periodicity of \(U\). Since
\begin{equation}
	U(x+2\pi,y)=U(x,y), \qquad U(x,y+2\pi)=U(x,y),
\end{equation}
we have
\begin{equation}
	P_{x+2\pi,y}(s) = s+\nu U(x+2\pi+s,y+\eta s) = P_{x,y}(s),
\end{equation}
and similarly
\begin{equation}
	P_{x,y+2\pi}(s) = s+\nu U(x+s,y+2\pi+\eta s) = P_{x,y}(s).
\end{equation}
Therefore, provided the zero is unique, one obtains
\begin{equation}
	\psi(x+2\pi,y)=\psi(x,y), \qquad \psi(x,y+2\pi)=\psi(x,y).
\end{equation}

The uniqueness of the zero follows when \(P_{x,y}(s)\) is monotonic. Indeed,
\begin{equation}
	\frac{d}{ds}P_{x,y}(s) = 1+\nu (\partial_x+\eta\partial_y) U(x+s,y+\eta s) = 1+\nu\left[ G(x+s,y+\eta s)-\nu^{-1} \right] = \nu G(x+s,y+\eta s).
\end{equation}
In the running regime, \(F(x,y)>0\), and hence
\begin{equation}
	G(x,y)=\frac1{F(x,y)}>0.
\end{equation}
Thus,
\begin{equation}
	\frac{d}{ds}P_{x,y}(s)>0,
\end{equation}
so \(P_{x,y}(s)\) has a unique zero. This proves that \(\psi(x,y)\) is a well-defined two-dimensional periodic function.

Consequently, the phase dynamics can be written as
\begin{equation}
	\boxed{
		\theta(t) = \nu t+\psi(\nu t,\eta\nu t)
	}.
\end{equation}
For irrational \(\eta\), the two phases \(\nu t\) and \(\eta\nu t\) wind incommensurately on the torus, producing quasiperiodic dynamics.

For numerical calculations, it is convenient to determine \(\psi(x,y)\) directly from the differential equation obtained by substituting
\[
\theta(t)=\nu t+\psi(\nu t,\eta\nu t)
\]
into the phase equation. Setting
\[
x=\nu t,\qquad y=\eta\nu t,
\]
one obtains
\begin{equation}
	\dot\theta = \nu \left( 1+\partial_x\psi+\eta\partial_y\psi \right).
\end{equation}
On the other hand,
\begin{equation}
	\dot\theta = F(x+\psi,y+\eta\psi).
\end{equation}
Therefore, \(\psi(x,y)\) satisfies the nonlinear first-order equation
\begin{equation}
	\boxed{
		\nu \left(1+\partial_x\psi+\eta\partial_y\psi\right) = F(x+\psi,y+\eta\psi)
	}.
\end{equation}
Equivalently,
\begin{equation}
	\partial_x\psi+\eta\partial_y\psi = \frac{1}{\nu} F(x+\psi,y+\eta\psi)-1 .
\end{equation}
For the present model,
\begin{equation}
	F(x,y) = \omega_\phi - A_1\cos(x+\varphi_1) - A_2\cos(y+\varphi_2),
\end{equation}
so that
\begin{equation}
	\boxed{
		\nu \left( 1+\partial_x\psi+\eta\partial_y\psi \right) = \omega_\phi - A_1\cos(x+\psi+\varphi_1)- A_2\cos(y+\eta\psi+\varphi_2)
	}.
\end{equation}
This equation is solved on the two-dimensional torus with periodic boundary conditions,
\begin{equation}
	\psi(x+2\pi,y)=\psi(x,y), \qquad \psi(x,y+2\pi)=\psi(x,y).
\end{equation}
To remove the gauge freedom associated with an overall phase shift, we fix the reference point by imposing
\begin{equation}
	\psi(0,0)=0 .
\end{equation}

\subsection{Quasiperiodic observables and frequency spectrum}

Any observable depending on the phase can be written as
\begin{equation}
	Q(t) = Q(\theta(t)).
\end{equation}
Using
\begin{equation}
	\theta(t) = \nu t+\psi(\nu t,\eta\nu t),
\end{equation}
we may equivalently regard \(Q\) as a periodic function on the torus,
\begin{equation}
	Q(t) = \mathcal Q(\nu t,\eta\nu t),
\end{equation}
where
\begin{equation}
	\mathcal Q(x,y) = Q(x+\psi(x,y)).
\end{equation}

Since \(\mathcal Q(x,y)\) is periodic in both arguments, it admits the Fourier expansion
\begin{equation}
	\mathcal Q(x,y) = \sum_{m,n\in\mathbb Z} Q_{mn}e^{i(mx+ny)},
\end{equation}
with
\begin{equation}
	Q_{mn} = \frac{1}{(2\pi)^2} \int_0^{2\pi}\int_0^{2\pi} \mathcal Q(x,y)e^{-i(mx+ny)}dx\,dy .
\end{equation}

Substituting
\begin{equation}
	x=\nu t, \qquad y=\eta\nu t,
\end{equation}
gives
\begin{equation}
	Q(t) = \sum_{m,n\in\mathbb Z} Q_{mn} \upe^{i(m+n\eta)\nu t}.
\end{equation}
Therefore, the frequency spectrum is supported on
\begin{equation}
	\boxed{
		\Omega_{mn} = (m+n\eta)\nu,\qquad m,n\in\mathbb Z.
	}
\end{equation}

For irrational \(\eta\), the frequencies \(\Omega_{mn}\) form a dense set on the real axis. The dynamics is
therefore quasiperiodic rather than periodic.

\subsection{Evaluation of the mean drift frequency}

The mean drift frequency is determined by the torus average
\begin{equation}
	\nu^{-1} = \bar g = \frac{1}{(2\pi)^2} \int_0^{2\pi}\int_0^{2\pi} \frac{dx\,dy}{\omega_\phi - A_1\cos x - A_2\cos y},
\end{equation}
where
\begin{equation}
	\omega_\phi>|A_1|+|A_2|
\end{equation}
is assumed so that the denominator remains strictly positive. Since the integral is invariant under translations of \(x\) and \(y\), the phase offsets \(\varphi_1\) and \(\varphi_2\) do not affect the final result and have been omitted.

We first perform the integration over \(x\). For fixed \(y\), define
\begin{equation}
	\beta = \omega_\phi-A_2\cos y .
\end{equation}
Then
\begin{equation}
	\bar g = \frac1{2\pi} \int_0^{2\pi} dy \left[\frac1{2\pi}\int_0^{2\pi}\frac{dx}{\beta-A_1\cos x}\right].
\end{equation}

Using the contour-integral substitution
\begin{equation}
	z=e^{\upi x}, \qquad \cos x=\frac12\left(z+\frac1z\right),
\end{equation}
the inner integral becomes
\begin{equation}
	\frac1{2\pi} \int_0^{2\pi} \frac{dx}{\beta-A_1\cos x} = \frac{1}{2\pi\upi} \oint_{|z|=1} \frac{2\,dz}{-A_1z^2+2\beta z-A_1}.
\end{equation}
The poles are located at
\begin{equation}
	z_\pm = \frac{\beta}{A_1} \pm\sqrt{\frac{\beta^2}{A_1^2}-1}.
\end{equation}
Since \(\beta>|A_1|\), only the pole \(z_-\) lies inside the unit circle. Evaluating the residue gives
\begin{equation}
	\frac1{2\pi} \int_0^{2\pi}\frac{dx}{\beta-A_1\cos x} = \frac1{\sqrt{\beta^2-A_1^2}} .
\end{equation}
Therefore,
\begin{equation}
	\bar g = \frac1{2\pi} \int_0^{2\pi}\frac{dy}{\sqrt{(\omega_\phi-A_2\cos y)^2-A_1^2}}.
\end{equation}

Next, we factorize the denominator,
\begin{equation}
	(\omega_\phi-A_2\cos y)^2-A_1^2=(\omega_\phi-A_1-A_2\cos y)(\omega_\phi+A_1-A_2\cos y).
\end{equation}
Introducing the half-angle substitution
\begin{equation}
	y=2x,
\end{equation}
together with
\begin{equation}
	\cos y = 1-2\sin^2x,
\end{equation}
yields
\begin{equation}
	\bar g =\frac1\pi \int_0^\pi\frac{dx}{\sqrt{(\omega_\phi-A_1-A_2+2A_2\sin^2x)(\omega_\phi+A_1-A_2+2A_2\sin^2x)}}.
\end{equation}

Using the symmetry about \(x=\pi/2\),
\begin{equation}
	\int_0^\pi f(\sin^2x)\,dx=2\int_0^{\pi/2}f(\sin^2x)\,dx ,
\end{equation}
we obtain
\begin{equation}
	\bar g = \frac2\pi \int_0^{\pi/2}\frac{dx}{\sqrt{(\alpha+\beta\sin^2x)(\gamma+\beta\sin^2x)}},
\end{equation}
where
\begin{equation}
	\alpha=\omega_\phi-A_1-A_2, \qquad \gamma=\omega_\phi+A_1-A_2, \qquad \beta=2A_2 .
\end{equation}

We now introduce the tangent substitution
\begin{equation}
	t=\tan x,
\end{equation}
so that
\begin{equation}
	\sin^2x=\frac{t^2}{1+t^2}, \qquad dx=\frac{dt}{1+t^2}.
\end{equation}
After straightforward algebra, the integral becomes
\begin{equation}
	\bar g = \frac2\pi \int_0^\infty \frac{dt}{\sqrt{(t^2+\alpha+\beta)(t^2+\gamma+\beta)}}.
\end{equation}

Next, we perform the rescaling
\begin{equation}
	t = \sqrt{\frac{\alpha}{\alpha+\beta}}\tan\theta .
\end{equation}
This transforms the integral into the standard Legendre form
\begin{equation}
	\bar g = \frac{2}{\pi\sqrt{(\alpha+\beta)\gamma}} \int_0^{\pi/2}\frac{d\theta}{\sqrt{1-k^2\sin^2\theta}},
\end{equation}
where
\begin{equation}
	k^2=\frac{\beta(\gamma-\alpha)}{\gamma(\alpha+\beta)} .
\end{equation}
Substituting the definitions of \(\alpha\), \(\beta\), and \(\gamma\),
one finds
\begin{equation}
	k^2=\frac{4A_1A_2}{(\omega_\phi+A_1-A_2)(\omega_\phi-A_1+A_2)}.
\end{equation}
Equivalently,
\begin{equation}
	k^2=\frac{4A_1A_2}{\omega_\phi^2-(A_1-A_2)^2}.
\end{equation}

The remaining integral is the complete elliptic integral of the first
kind,
\begin{equation}
	K(k)=\int_0^{\pi/2}\frac{d\theta}{\sqrt{1-k^2\sin^2\theta}}.
\end{equation}
Therefore,
\begin{equation}
	\boxed{
		\bar g = \frac{2}{\pi\sqrt{\omega_\phi^2-(A_1-A_2)^2}}K(k).
	}
\end{equation}
Since
\begin{equation}
	\omega_\phi>|A_1|+|A_2|,
\end{equation}
one has
\begin{equation}
	\omega_\phi^2-(A_1-A_2)^2>(A_1+A_2)^2-(A_1-A_2)^2=4A_1A_2,
\end{equation}
and therefore
\begin{equation}
	0\leq k^2<1 .
\end{equation}

Finally, the mean drift frequency is
\begin{equation}
	\boxed{
		\nu = \bar g^{-1} = \frac{\pi\sqrt{\omega_\phi^2-(A_1-A_2)^2}}{2K(k)}.
	}
\end{equation}

In the limit \(A_2\to0\), one has \(k\to0\) and
\begin{equation}
	K(0)=\frac{\pi}{2}.
\end{equation}
The result then reduces to
\begin{equation}
	\nu = \sqrt{\omega_\phi^2-A_1^2},
\end{equation}
which reproduces the standard two-level boundary time-crystal result.

\bibliography{Refs_SM}

\end{document}